\documentclass[12pt]{article}
\pdfoutput=1

\usepackage{graphicx}
\usepackage{amsfonts}
\usepackage{amssymb}
\usepackage{amsthm}
\usepackage{amsmath}

\usepackage[active]{srcltx}

\usepackage{cite}
\usepackage{color,soul}

\usepackage[affil-it]{authblk}

\graphicspath{{figs/}}

                                   \usepackage{ulem}

\newcommand{\be}{\begin{equation}}
\newcommand{\ee}{\end{equation}}

\allowdisplaybreaks
\newcommand{\lsim}
{\;\raisebox{-.3em}{$\stackrel{\displaystyle <}{\sim}$}\;}

\newcommand{\SM}{{\rm SM}}

\newcommand\al{\alpha}
\newcommand\bet{\beta}
\newcommand\tb{\tan\beta}

\newcommand\ReDiag{\mathop{%
  \raise .5pt\hbox{[}%
  \widetilde{\mathrm{Re}}%
  \raise .5pt\hbox{]}}}
\newcommand\ReOffDiag{\mathop{%
  \raise .5pt\hbox{$\llbracket$}%
  \widetilde{\mathrm{Re}}%
  \raise .5pt\hbox{$\rrbracket$}}}

\newcommand\MZ{M_Z}
\newcommand\Mh{M_h}
\newcommand\MH{M_H}
\newcommand\MA{M_A}
\newcommand\MHp{M_{H^\pm}}

\newcommand\mb{m_b}
\newcommand\mt{m_t}

\newcommand\gl{{\tilde g}}
\newcommand\mgl{m_\gl}

\newcommand\ino[1]{\tilde\chi_{#1}}

\newcommand\chapm[1]{\ino{#1}^\pm}

\newcommand\mcha[1]{m_{\chapm{#1}}}

\newcommand\neu[1]{\ino{#1}^0}
\newcommand\mneu[1]{m_{\neu{#1}}}

\newcommand\refeq[1]{Eq.~(\ref{#1})}
\newcommand\refeqs[1]{Eqs.~(\ref{#1})}
\newcommand\refta[1]{Tab.~\ref{#1}}
\newcommand\refse[1]{Sect.~\ref{#1}}
\newcommand\refses[1]{Sects.~\ref{#1}}
\newcommand\citere[1]{Ref.~\cite{#1}}
\newcommand\citeres[1]{Refs.~\cite{#1}}

\newcommand{\CP}{{\cal CP}}
\newcommand{\cp}{{\CP}}

\newcommand{\tev}{\,\, \mathrm{TeV}}
\newcommand{\gev}{\,\, \mathrm{GeV}}

\newcommand\FH{\texttt{FeynHiggs}}

\newcommand\mstop[1]{m_{\tilde{t}_{#1}}}
\newcommand\msbot[1]{m_{\tilde{b}_{#1}}}

\newcommand\mstau[1]{m_{\tilde{\tau}_{#1}}}

\newcommand{\br}{\text{BR}}

\newcommand{\sig}{\sigma}

\def\order#1{\ensuremath{{\cal O}(#1)}}
\def\reffi#1{\mbox{Fig.~\ref{#1}}}

\def\ga{\gamma}
\def\De{\Delta}
\def\de{\delta}

\newcommand{\VL}{\left( \begin{array}{c}}
\newcommand{\VR}{\end{array} \right)}
\newcommand{\ML}{\left( \begin{array}{cc}}
\newcommand{\MLd}{\left( \begin{array}{ccc}}
\newcommand{\MLv}{\left( \begin{array}{cccc}}
\newcommand{\MR}{\end{array} \right)}

\definecolor{Lightblue}{cmyk}{0.9,0.1,0.1,0.3}
\definecolor{dgelborange}{cmyk}{0.,0.3,0.5, 0.}
\definecolor{Lila}{rgb}{0.5,0.,1}

\begin{document}

\title{
\vspace*{-3cm}
\hfill {\small CERN-TH-2017-225}\\[-0.7em]
\hfill {\small IFT-UAM/CSIC-17-116}\\
\vspace*{1cm}
Reduction of the Parameters in MSSM}
\date{}
\author{S. Heinemeyer$^{1,2,3}$\thanks{email: Sven.Heinemeyer@cern.ch} , M. Mondrag\'on$^4$\thanks{email: myriam@fisica.unam.mx} , N. Tracas$^{5,6}$\thanks{email: ntrac@central.ntua.gr}$\,$ and G. Zoupanos$^{5,6,7}$\thanks{email: George.Zoupanos@cern.ch}}

\affil{\small

$^1$Instituto de F\'{\i}sica Te\'{o}rica (UAM/CSIC), Universidad Auton\'{o}ma de Madrid, 28049 Madrid, Spain\\
$^2$Campus of International Excellence UAM+CSIC,
Cantoblanco, 28049, Madrid, Spain\\
$^3$Instituto de F\'{\i}sica de Cantabria (CSIC-UC),  E-39005 Santander, Spain \\
$^4$Instituto de F\'{\i}sica,  Universidad Nacional
Aut\'onoma de M\'exico,  A.P. 20-364, CDMX 01000, M\'exico\\ %
$^5$ Theoretical Physics Department, CERN, Geneva, Switzerland\\%
$^6$ Physics Department, Nat.\ Technical University, 157 80 Zografou, Athens,
Greece\\
$^7$Max-Planck Institut f\"ur Physik, F\"ohringer Ring 6, D-80805
  M\"unchen, Germany
}

\maketitle

\abstract{In the present work we search for renormalization group invariant
  relations among the various massless and massive parameters of the Minimal
  Supersymmetric Standard Model. We find that indeed several of the previously
  free parameters of the model can be reduced in favor of few, the unique
  gauge coupling and the gaugino mass at the unification scale among
  them. Taking into account the various experimental constraints, including
  the $B$-physics ones, we predict the Higgs and the supersymmetric spectrum. We
  find that the lightest Higgs mass is in comfortable agreement with the
  measured value and its experimental and theoretical uncertainties, while the
  electroweak supersymmetric spectrum starts at 1.3~TeV and the colored at
  $\sim$4~TeV. Thus the reduced MSSM is in natural agreement with all LHC
  measurements and searches. The supersymmetric and heavy Higgs particles will
  likely escape the detection at the LHC, as well as at ILC and CLIC. However,
  the FCC-hh will be able to fully test the predicted parameter space.
}

\vfill
\newpage

%%%%%%%%%%%%%%%%%%%%%%%%%%%%%%%%%%%%%%%%%%%%%%%%%%%%%%%%%%%%%%%%%%%%%%%%%%%%%%%
%%%%%%%%%%%%%%%%%%%%%%%%%%%%%%%%%%%%%%%%%%%%%%%%%%%%%%%%%%%%%%%%%%%%%%%%%%%%%%%

\section{Introduction}

The main expectation of the particle physics community from a unified description of the observed interactions is
to understand the present day large number of free parameters of the Standard Model (SM) in terms of a few fundamental ones.
In other words, to achieve \textit{reduction of parameters} at a
fundamental level.

The traditional way to reduce the number of free parameters of a theory, which in turn would make it more predictive,
is to introduce a symmetry. Grand Unified Theories (GUTs) are very good examples of this strategy
\cite{Pati:1973rp,Georgi:1974sy,Georgi:1974yf,Fritzsch:1974nn,Carlson:1975gu}.
In the case of minimal $SU(5)$, because of the (approximate) gauge coupling unification, it was possible to reduce the gauge couplings of the
SM and give a prediction for one of them. In fact, the LEP data
\cite{Amaldi:1991cn}
were interpreted as suggesting that a further symmetry, namely $N = 1$ global supersymmetry (SUSY)
\cite{Dimopoulos:1981zb,Sakai:1981gr}
should also be required to make the prediction viable. GUTs can also relate the Yukawa couplings among
themselves, again $SU(5)$ provided an example of this by predicting the ratio $M_\tau/M_b$
\cite{Buras:1977yy}
in the SM.
Unfortunately, requiring more symmetry does not necessarily helps, since additional complications are introduced due to new degrees
of freedom that normally are needed, requiring in turn new ways and channels of breaking the symmetry, among others, which in general reduce the predictivity of a theory.

A natural extension of the GUT idea is to find a way to relate the gauge and Yukawa sectors of a theory, that is to
achieve Gauge-Yukawa Unification (GYU). A symmetry which naturally relates the two sectors is SUSY, in particular $N = 2$ SUSY
\cite{Fayet:1978ig}.
However, $N = 2$ supersymmetric theories have serious phenomenological problems due to light mirror fermions. Other theories
such as superstring theories or composite models might provide relations among the gauge and Yukawa couplings, but have even
more phenomenological problems. A successful strategy in relating dimensionless couplings has been developed in a series of studies
\cite{Kubo:1995cg,Kubo:1997fi,Kobayashi:1999pn,Kapetanakis:1992vx,Mondragon:1993tw,Kubo:1994bj,Kubo:1994xa,Kubo:1995zg,Kubo:1996js,MMZ:04,MTZ:14}.
It was based on searches for renormalization group invariant (RGI) relations. This program, called Gauge-Yukawa unification scheme,
applied in the dimensionless couplings of supersymmetric GUTs, such as gauge and Yukawa couplings, had already celebrated successes
by predicting correctly, among others, the top quark mass in the finite and in the minimal $N = 1$ supersymmetric $SU(5)$ GUTs
\cite{Kapetanakis:1992vx,Mondragon:1993tw,Kubo:1994bj},
$SU(3)^3$
\cite{MMZ:04}
and later in the Minimal Supersymmetric Standard Model (MSSM)
\cite{MTZ:14}.
One of the impressive aspects of the RGI relations is that their validity can be guaranteed to all-orders in perturbation
theory by studying the uniqueness of the resulting relations at one-loop, as was proven
\cite{Zimmermann:1984sx,Oehme:1984yy}
in the early days of the program of reduction of couplings
\cite{Zimmermann:1984sx,Oehme:1984yy,Ma:1977hf,Ma:1984by,Chang:1974bv,Nandi:1978fw}.
Even more impressive is the fact that it is possible to find RGI relations among couplings guaranteeing finiteness
to all-orders in perturbation theory
\cite{Lucchesi:1987ef,Lucchesi:1987he,Lucchesi:1996ir,Ermushev:1986cu,Kazakov:1987vg}.

SUSY seems to be an essential ingredient for a plenomenologically successful realization of the above strategy.
Nevertheless its breaking has to be understood too in order to extend the successes in other sectors of the theory,
such as the Higgs masses and the SUSY spectrum.

Indeed, the search for RGI relations has been extended to the soft SUSY-breaking sector (SSB) of these theories
\cite{Kubo:1996js,Jack:1995gm},
which involves parameters of dimension one and two. The first important development in this programme concerned the combined
reduction of couplings and masses in supersymmetric theories
\cite{Kubo:1996js}.
In this work the coefficients of the soft SUSY-breaking terms were reduced in order to minimize the number of independent
parameters. The scheme of dimensional renormalization was used with mass parameters introduced similarly to couplings. Then the
differential equations of the renormalization group also involve derivatives with respect to the masses. It is characteristic
for dimensional renormalization that those $\beta$-functions which carry a dimension are linear or quadratic forms in the dimensional
couplings and masses, while the coefficients of these polynomials depend on the dimensionless couplings only. Since in this
approach the mass parameters enter similarly to the couplings, masses are included with the couplings in the reduction process.
In this way non-trivial constraints on the soft SUSY-breaking terms were obtained which are compatible with renormalization
and lead to surprisingly simple sum rules
\cite{Kawamura:1997cw}.

Another very important development concerning the renormalization
properties of the SSB was made in
\citeres{Hisano:1997ua,Jack:1997pa,Avdeev:1997vx,Kazakov:1998uj,Kazakov:1997nf,Jack:1997eh,Kobayashi:1998jq},
based conceptually and technically on the work of Ref.
\cite{Yamada:1994id}:
the powerful supergraph method
\cite{Delbourgo:1974jg,Salam:1974pp,Fujikawa:1974ay,Grisaru:1979wc}
for studying supersymmetric theories was applied to the softly broken ones by using the ``spurion'' external space-time independent superfields
\cite{Girardello:1981wz}.
In the latter method a softly broken supersymmetric gauge theory is considered as a supersymmetric one
in which the various parameters such as couplings and masses have been promoted to external superfields that acquire ``vacuum expectation
values''. Based on this method the relations among the soft term renormalization and that of an unbroken supersymmetric theory were derived.
In particular the $\beta$-functions of the parameters of the softly broken theory are expressed in terms of partial differential operators
involving the dimensionless parameters of the unbroken theory. The key point in the
strategy of Refs.
\cite{Kazakov:1998uj,Kazakov:1997nf,Jack:1997eh,Kobayashi:1998jq}
in solving the set of coupled differential equations so as to be able to express all parameters in a RGI way, was to transform the partial differential operators involved to total derivative operators. This is indeed possible to be done on the RGI surface
which is defined by the solution of the reduction equations. The last has very important consequences in the finite theories since the finiteness of the dimensionless sector can be transferred to the SSB sector too.

In parallel to the above theoretical developments certain phenomenological issues have been established too.
For long time a rather constrained universal set of soft scalar masses has been assumed in the SSB sector of supersymmetric theories not only
for economy and simplicity but for a number of other reason too: (a) they were part of the constraints that preserve finiteness up to two-loops
\cite{Jones:1984cu,Jack:1994kd},
(b) they are RGI up to two-loops in more general supersymmetric gauge theories, subject to
the condition known as $P = 1/3 Q$~\cite{Jack:1995gm} (where all relevant
details and definitions can be found),
and (c) they appear in the attractive dilaton dominated SUSY-breaking superstring scenarios
\cite{Ibanez:1992hc,Kaplunovsky:1993rd,Brignole:1993dj}.
However, further studies have shown that there exist a number of technical problems all due to the fact that the
universality assumption for the soft scalar masses is very restrictive. For instance, (i) in finite unified theories
the universality predicts that the lightest supersymmetric  particle is a charged particle, namely the superpartner of
the $\tau$-lepton, (ii) the standard radiative electroweak symmetry breaking of the MSSM
does not work with universal soft scalar masses
\cite{Brignole:1993dj},
and (iii) which is more serious, the universal soft scalar masses lead to charge and/or color breaking minima deeper
than the standard vacuum \cite{Casas:1996wj}.
In addition criticisms arose on an aesthetic basis, i.e. that the
universal assumption is too strong to be put by hand given that it
does not result from something fundamental. A way out was indirectly already suggested in ref
\cite{Kubo:1996js},
where the solutions found among soft scalar masses were very different from the universal one. Moreover a more careful look
suggested the existence of a ``sum rule'' among the soft scalar masses and the gaugino mass. This interesting observation was clearly done in ref
\cite{Kawamura:1997cw}
where it was examined in $N = 1$ Gauge-Yukawa unified theories at one-loop for the non-finite case and then at two-loops for the finite case
\cite{Kobayashi:1997qx}.
The sum rule manages to overcome all the unpleasant phenomenological consequences mentioned above. Moreover it was proven
\cite{Kobayashi:1998jq}
that the sum rule for the soft scalar masses is RGI to all-orders for both the general as well as for the finite case.
Finally, the exact $\bet$-function for the soft scalar masses in the Novikov-Shifman-Vainstein-Zakharov (NSVZ) scheme
\cite{Novikov:1983ee,Novikov:1985rd,Shifman:1996iy}
for the softly broken supersymmetric QCD has been obtained
\cite{Kobayashi:1998jq}.

Using the above tools and results it was possible to study and predict the spectrum of the full finite models in terms of few input parameters.
A particular finite model was selected out of this examination and provided us with the prediction for the lightest MSSM Higgs boson in the range of
121-126~Gev
\cite{finite,finite_2,finite_3,finite_4}
four and half years before the experimental discovery
\cite{:2012gk,:2012gu}.%
\footnote{It should be kept in mind that this prediction did not
    yet include the resummation of large logarithmic contributions to
    the light Higgs boson mass, see the discussion in
    \refse{sec:constraints}.}%
~Identifying the lightest Higgs boson with the newly discovered state one can restrict the allowed parameter space of the model.
A similar analysis was done for the reduced MSSM
\cite{MTZ:14}.

In the present work we examine the reduced MSSM using the ``exact'' relations among soft scalar and gaugino masses, following the original analysis suggested in ref
\cite{Kubo:1996js}.
Obviously the reduced MSSM in the present case is much more constrained as compared to the previous one
\cite{MTZ:14},
which was enjoying the benefit of the relaxed ``sum rule''.
The results are confronted with the relevant flavor physics
results. We evaluate the full SUSY spectrum (for sfermions restricted to
the third generation), which turns out to be
rather heavy, and in particular we calculate the
lightest MSSM Higgs-boson mass. Here, in contrast to previous
evaluations, an improved calculation is employed that yields more
reliable results for heavy SUSY masses. The light Higgs-boson mass is
naturally found in the region of $124 - 129 \gev$.

%%%%%%%%%%%%%%%%%%%%%%%%%%%%%%%%%%%%%%%%%%%%%%%%%%%%%%%%%%%%%%%%%%%%%%%%%%%%%%%
%%%%%%%%%%%%%%%%%%%%%%%%%%%%%%%%%%%%%%%%%%%%%%%%%%%%%%%%%%%%%%%%%%%%%%%%%%%%%%%

\section{Reduction of Parameters}

The reduction of couplings was originally formulated for massless theories on the basis of
the Callan-Symanzik equation
\cite{Zimmermann:1984sx,Oehme:1984yy}.
The extension to theories with massive parameters is not
straightforward if one wants to keep the generality and the rigor on the same level as for the
massless case; one has to fulfill a set of requirements coming from the renormalization group
equations, the Callan-Symanzik equations, etc. along with the normalization conditions
imposed on irreducible Green's functions
\cite{piguet1}.
There has been a lot of progress in this direction starting from ref. \cite{Kubo:1996js},
as it is already mentioned in the Introduction, where it was assumed that a mass-independent
renormalization scheme could be employed so that all the RG functions have only trivial
dependencies on dimensional parameters and then the mass parameters were
introduced similarly to couplings (i.e.\ as a power series in the
couplings). This choice was
justified later in \cite{zim2,Zimer:2001n} where
the scheme independence of the reduction principle has been proven generally, i.e\ it was shown that apart from dimensionless couplings,
pole masses and gauge parameters, the model may also involve coupling parameters carrying a dimension and masses.
Therefore here, to simplify the analysis, we follow \citere{Kubo:1996js} and
we too use a mass-independent renormalization scheme.

We start by considering a renormalizable theory which contain a set of $(N + 1)$
dimension-zero couplings, $\left(\hat g_0,\hat g_1, ...,\hat g_N\right)$,
a set of $L$ parameters with mass-dimension one, $\left(\hat h_1,...,\hat h_L\right)$,
and a set of $M$ parameters with mass-dimension two, $\left(\hat m_1^2,...,\hat m_M^2\right)$.
The renormalized irreducible vertex function $\Gamma$ satisfies the RG
equation
\be
\label{RGE_OR_1}
\mathcal{D}\Gamma\left[\Phi's;\hat g_0,\hat g_1, ...,\hat g_N;\hat h_1,...,\hat h_L;\hat m_1^2,...,\hat m_M^2;\mu\right]=0~,
\ee
where
\be
\label{RGE_OR_2}
\mathcal{D}=\mu\frac{\partial}{\partial \mu}+
\sum_{i=0}^N \beta_i\frac{\partial}{\partial \hat g_i}+
\sum_{a=1}^L \gamma_a^h\frac{\partial}{\partial \hat h_a}+
\sum_{\alpha=1}^M \gamma_\alpha^{m^2}\frac{\partial}{\partial \hat m_\alpha ^2}+
\sum_J \Phi_I\gamma^{\phi I}_{\,\,\,\, J}\,\frac{\delta}{\delta\Phi_J}~,
\ee
where $\mu$ is the energy scale,
while $\beta_i$ are the $\beta$-functions of the various dimensionless
couplings $g_i$, $\Phi_I$  are
the various matter fields and
$\ga_\alpha^{m^2}$, $\ga_a^h$ and $\ga^{\phi I}_{\,\,\,\, J}$
are the mass, trilinear coupling and wave function anomalous dimensions,
respectively
(where $I$ enumerates the matter fields).
In a mass independent renormalization scheme, the $\gamma$'s are given by
\be
\label{gammas}
\begin{split}
\gamma^h_a&=\sum_{b=1}^L\gamma_a^{h,b}(g_0,g_1,...,g_N)\hat h_b,\\
\gamma_\alpha^{m^2}&=\sum_{\beta=1}^M \gamma_\alpha^{m^2,\beta}(g_0,g_1,...,g_N)\hat m_\beta^2+
\sum_{a,b=1}^L \gamma_\alpha^{m^2,ab}(g_0,g_1,...,g_N)\hat h_a\hat h_b,
\end{split}
\ee
where $\gamma_a^{h,b}$, $\gamma_\alpha^{m^2,\beta}$ and $\gamma_\alpha^{m^2,ab}$ are power series of the
$g$'s (which are dimensionless) in perturbation theory.\\

We look for a reduced theory where
\[
g\equiv g_0,\qquad h_a\equiv \hat h_a\quad \textrm{for $1\leq a\leq P$},\qquad
m^2_\alpha\equiv\hat m^2_\alpha\quad \textrm{for $1\leq \alpha\leq Q$}
\]
are independent parameters and the reduction of the parameters left
\be
\label{reduction}
\begin{split}
\hat g_i &= \hat g_i(g), \qquad (i=1,...,N),\\
\hat h_a &= \sum_{b=1}^P f_a^b(g)h_b, \qquad (a=P+1,...,L),\\
\hat m^2_\alpha &= \sum_{\beta=1}^Q e^\beta_\alpha(g)m^2_\beta + \sum_{a,b=1}^P k^{ab}_\alpha(g)h_ah_b,
\qquad (\alpha=Q+1,...,M)
\end{split}
\ee
is consistent with the RG equations (\ref{RGE_OR_1},\ref{RGE_OR_2}). It
turns out that the following relations should be satisfied
\be
\label{relation}
\begin{split}
\beta_g\,\frac{\partial\hat g_i}{\partial g} &= \beta_i,\qquad (i=1,...,N),\\
\beta_g\,\frac{\partial \hat h_a}{\partial g}+\sum_{b=1}^P \gamma^h_b\,\frac{\partial\hat h_a}{\partial h_b} &= \gamma^h_a,\qquad (a=P+1,...,L),\\
\beta_g\,\frac{\partial\hat m^2_\alpha}{\partial g} +\sum_{a=1}^P \gamma_a^h\,\frac{\partial\hat m^2_\alpha}{\partial h_a} +  \sum_{\beta=1}^Q \gamma_\beta ^{m^2}\,\frac{\partial\hat m_\alpha^2}{\partial m_\beta^2} &= \gamma_\alpha^{m^2}, \qquad (\alpha=Q+1,...,M).
\end{split}
\ee
Using \refeqs{gammas} and (\ref{reduction}), the above relations reduce to
\be
\label{relation_2}
\begin{split}
&\beta_g\,\frac{df^b_a}{dg}+ \sum_{c=1}^P f^c_a\left[\gamma^{h,b}_c + \sum_{d=P+1}^L \gamma^{h,d}_c f^b_d\right] -\gamma^{h,b}_a - \sum_{d=P+1}^L \gamma^{h,d}_a f^b_d=0,\\
&(a=P+1,...,L;\, b=1,...,P),\\
&\beta_g\,\frac{de^\beta_\alpha}{dg} + \sum_{\gamma=1}^Q e^\gamma_\alpha\left[\gamma_\gamma^{m^2,\beta} +
\sum _{\delta=Q+1}^M\gamma_\gamma^{m^2,\delta} e^\beta_\delta\right]-\gamma_\alpha^{m^2,\beta} -
\sum_{\delta=Q+1}^M \gamma_\alpha^{m^2,d}e^\beta_\delta =0,\\
&(\alpha=Q+1,...,Mq\, \beta=1,...,Q),\\
&\beta_g\,\frac{dk_\alpha^{ab}}{dg}
+ 2\sum_{c=1}^P \left(\gamma_c^{h,a} + \sum_{d=P+1}^L \gamma_c^{h,d} f_d^a\right)k_\alpha^{cb}
+\sum_{\beta=1}^Q e^\beta_\alpha\left[\gamma_\beta^{m^2,ab} + \sum_{c,d=P+1}^L \gamma_\beta^{m^2,cd}f^a_cf^b_d \right.\\
&\left. +2\sum_{c=P+1}^L \gamma_\beta^{m^2,cb}f^a_c + \sum_{\delta=Q+1}^M \gamma_\beta^{m^2,d} k_\delta^{ab}\right]- \left[\gamma_\alpha^{m^2,ab}+\sum_{c,d=P+1}^L \gamma_\alpha^{m^2,cd}f^a_c f^b_d\right.\\
&\left. +2 \sum_{c=P+1}^L \gamma_\alpha^{m^2,cb}f^a_c + \sum_{\delta=Q+1}^M \gamma_\alpha^{m^2,\delta}k_\delta^{ab}\right]=0,\\
&(\alpha=Q+1,...,M;\, a,b=1,...,P)~.
\end{split}
\ee
The above relations ensure that the irreducible vertex function of the reduced theory
\be
\label{Green}
\begin{split}
\Gamma_R&\left[\Phi\textrm{'s};g;h_1,...,h_P; m_1^2,...,m_Q^2;\mu\right]\equiv\\
&\Gamma \left[  \Phi\textrm{'s};g,\hat g_1(g)...,\hat g_N(g);
h_1,...,h_P,\hat h_{P+1}(g,h),...,\hat h_L(g,h);\right.\\
& \left.  \qquad\qquad\qquad m_1^2,...,m^2_Q,\hat m^2_{Q+1}(g,h,m^2),...,\hat m^2_M(g,h,m^2);\mu\right]
\end{split}
\ee
has the same renormalization group flow as the original one.

The assumptions that the reduced theory is perturbatively renormalizable
means that the functions
$\hat g_i$, $f^b_a$, $e^\beta_\alpha$ and $k_\alpha^{ab}$, defined in (\ref{reduction}), should be
expressed as a power series in the primary coupling $g$:
\be
\label{pert}
\begin{split}
\hat g_i & = g\sum_{n=0}^\infty \rho_i^{(n)} g^n,\qquad
f_a^b  =  g \sum_{n=0}^\infty \eta_a^{b(n)} g^n\\
e^\beta_\alpha & = \sum_{n=0}^\infty \xi^{\beta(n)}_\alpha g^n,\qquad
k_\alpha^{ab}=\sum_{n=0}^\infty \chi_\alpha^{ab(n)} g^n.
\end{split}
\ee
The above expansion coefficients can be found by inserting these power
series into \refeqs{relation}, (\ref{relation_2}) and requiring the equations to be satisfied at each order of $g$.
It should be noted that the existence of a unique power series solution is a non-trivial matter: It depends on the theory
as well as on the choice of the set of independent parameters.

It should also be noted that in the case that there are no independent
mass-dimension~1 parameters ($\hat h$) the reduction of these terms take
naturally the form
\[
\hat h_a = \sum_{b=1}^L f_a^b(g)M,
\]
where $M$ is a mass-dimension 1 parameter which could be a gaugino mass which corresponds to the independent (gauge) coupling. In case, on top of that, there are no independent
mass-dimension 2 parameters ($\hat m^2$), the corresponding reduction takes analogous form
\[
\hat m^2_a=\sum_{b=1}^M e_a^b(g) M^2.
\]

%%%%%%%%%%%%%%%%%%%%%%%%%%%%%%%%%%%%%%%%%%%%%%%%%%%%%%%%%%%%%%%%%%%%%%%%%%%%%%%
%%%%%%%%%%%%%%%%%%%%%%%%%%%%%%%%%%%%%%%%%%%%%%%%%%%%%%%%%%%%%%%%%%%%%%%%%%%%%%%

\section{Reduction of dimensionless parameters in the MSSM}
\label{sec:red-dim0}

Hereafter we are working in the framework of MSSM, assuming though the existence of a covering GUT.
The superpotential of the MSSM (where again we restrict ourselves to the
  third generation of sfermions) is defined by
\be
\label{supot2}
W = Y_tH_2Qt^c+Y_bH_1Qb^c+Y_\tau H_1L\tau^c+ \mu H_1H_2\, ,
\ee
where $Q,L,t,b,\tau, H_1,H_2$ are the usual superfields of MSSM,
while the SSB Lagrangian is given by
\be
\label{SSB_L}
\begin{split}
-\mathcal{L}_{\rm SSB} &= \sum_\phi m^2_\phi\phi^*\phi+
\left[m^2_3H_1H_2+\sum_{i=1}^3 \frac 12 M_i\lambda_i\lambda_i +\textrm{h.c}\right]\\
&+\left[h_tH_2Qt^c+h_bH_1Qb^c+h_\tau H_1L\tau^c+\textrm{h.c.}\right] ,          %%%%%\qquad       (45)
\end{split}
\ee
where $\phi$ represents the scalar component of all superfields, $\lambda$ refers to the gaugino fields
while in the last brace we refer to the scalar components of the corresponding superfield.
The Yukawa $Y_{t,b,\tau}$ and the trilinear $h_{t,b,\tau}$ couplings  refer to the third generator only,
neglecting the first two generations.

Let us start with the dimensionless couplings, i.e. gauge and Yukawa. As a first step we consider
only the strong coupling and the top and bottom Yukawa couplings, while the other two gauge couplings and the tau
Yukawa will be treated as corrections.
Following the above line, we reduce the Yukawa couplings in favor of
the strong coupling~$\al_3$
\[
\frac{Y^2_i}{4\pi}\equiv \al_i=G_i^2\al_3,\qquad i=t,b,
\]
and using the RGE for the Yukawa, we get
\[
G_i^2=\frac 13 ,\qquad i=t,b.
\]
This system of the top and bottom Yukawa couplings reduced with the strong one
is dictated by (i) the different running behaviour of the
$SU(2)$ and $U(1)$ coupling compared to the strong one
\cite{KSZ_1985}
and (ii) the incompatibility of applying the above reduction
for the tau Yukawa since the corresponding $G^2$ turns negative
\cite{{MTZ:14}}.
Adding now the two other gauge couplings and the tau Yukawa in the RGE as
corrections, we obtain
\be
\label{Gt2_Gb2}
G_t^2=\frac 13+\frac{71}{525}\rho_1+\frac 37 \rho_2 +\frac 1{35}\rho_\tau,\qquad
G_b^2=\frac 13+\frac{29}{525}\rho_1+\frac 37 \rho_2 -\frac 6{35}\rho_\tau
\ee
where
\be
\label{r1_r2_rtau}
\rho_{1,2}=\frac{g_{1,2}^2}{g_3^2}=\frac{\al_{1,2}}{\al_3},\qquad
\rho_\tau=\frac{g_\tau^2}{g_3^2}=\frac{\displaystyle{\frac{Y^2_\tau}{4\pi}}}{\al_3}
\ee

Note that the corrections in Eq.(\ref{Gt2_Gb2}) are taken at the GUT scale and under the assumption that
\[
\frac{d}{dg_3}\left(\frac {Y_{t,b}^2}{g_3^2}\right)=0.
\]

Let us comment further on our assumption above, which led to the Eq.(\ref{Gt2_Gb2}).
In practice we assume that even including the corrections from the rest of the gauge as well as the tau Yukawa couplings,
at the GUT scale the ratio of the top and bottom couplings $\al_{t,b}$ over the strong coupling are still constant,
i.e. their scale dependence is negligible.
Or, rephrasing it, our assumption can be understood as a requirement that in the ultraviolet (close to the GUT scale)
the ratios of the top and bottom Yukawa couplings over the strong coupling become least sensitive against the change
of the renormalization scale. This requirement sets the boundary condition at the GUT scale, given in Eq.(\ref{Gt2_Gb2}).
Alternatively one could follow the systematic method to include the corrections to a non-trivially reduced system developed
in ref\cite{KSZ_1989}, but considering two reduced systems: the first one consisting of the ``top, bottom'' couplings and the second of the
``strong, bottom'' ones. We plan to return with the full analysis of the latter possibility, including the dimensionful parameters,
in a future publication.

In the next order the corrections are assumed to be in the form
\[
\al_i=G_i^2\al_3+J_i^2 \al_3^2,\qquad i=t,b.
\]
Then, the coefficients $J_i$ are given by
\[
J_i^2=\frac 1{4\pi}\,\frac{17}{24},\qquad i=t,b
\]
for the case where only the strong gauge and the top and bottom Yukawa couplings are active,
while for the case where the other two gauge and the tau Yukawa couplings are
added as corrections we obtain
\[
J_t^2=\frac 1{4\pi}\frac{N_t}{D},\quad
J_b^2=\frac 1{4\pi}\frac{N_b}{5D},
\]
where
\[
\begin{split}
D=&
257250 (196000 + 44500 \rho_1 + 2059 \rho_1^2 + 200250 \rho_2 + 22500 \rho_1 \rho_2 +
   50625 \rho_2^2 - \\
&33375 \rho_\tau - 5955 \rho_1 \rho_\tau - 16875 \rho_2 \rho_\tau -
   1350 \rho_\tau^2),
\end{split}
\]

\[
\begin{split}
N_t=&
-(-35714875000 - 10349167500 \rho_1 + 21077903700 \rho_1^2 +9057172327 \rho_1^3 +\\
&   481651575 \rho_1^4 - 55566000000 \rho_2 +2857680000 \rho_1 \rho_2 + 34588894725 \rho_1^2 \rho_2 +\\
&   5202716130 \rho_1^3 \rho_2 +3913875000 \rho_2^2 + 8104595625 \rho_1 \rho_2^2 + 11497621500 \rho_1^2 \rho_2^2 +\\
&   27047671875 \rho_2^3 + 1977918750 \rho_1 \rho_2^3 + 7802578125 \rho_2^4 +3678675000 \rho_\tau +\\
&   1269418500 \rho_1 \rho_\tau - 2827765710 \rho_1^2 \rho_\tau -1420498671 \rho_1^3 \rho_\tau +7557637500 \rho_2 \rho_\tau -\\
&   2378187000 \rho_1 \rho_2 \rho_\tau - 4066909425 \rho_1^2 \rho_2 \rho_\tau -1284018750 \rho_2^2 \rho_\tau - 1035973125 \rho_1 \rho_2^2 \rho_\tau -\\
&   2464171875 \rho_2^3 \rho_\tau + 1230757500 \rho_\tau^2 + 442136100 \rho_1 \rho_\tau^2 -186425070 \rho_1^2 \rho_\tau^2 +\\
&   1727460000 \rho_2 \rho_\tau^2 +794232000 \rho_1 \rho_2 \rho_\tau^2 + 973518750 \rho_2^2 \rho_\tau^2 -\\
&   325804500 \rho_\tau^3 - 126334800 \rho_1 \rho_\tau^3 - 412695000 \rho_2 \rho_\tau^3 -
   32724000 \rho_\tau^4),
\end{split}
\]

\[
\begin{split}
N_b=&
-(-178574375000 - 71734162500 \rho_1 + 36055498500 \rho_1^2 +13029194465 \rho_1^3 +\\
&   977219931 \rho_1^4 - 277830000000 \rho_2 -69523650000 \rho_1 \rho_2 + 72621383625 \rho_1^2 \rho_2 +\\
&   10648126350 \rho_1^3 \rho_2 +19569375000 \rho_2^2 + 13062459375 \rho_1 \rho_2^2 + 25279672500 \rho_1^2 \rho_2^2 +\\
&   135238359375 \rho_2^3 + 16587281250 \rho_1 \rho_2^3 + 39012890625 \rho_2^4 +58460062500 \rho_\tau +\\
&   35924411250 \rho_1 \rho_\tau - 13544261325 \rho_1^2 \rho_\tau -2152509435 \rho_1^3 \rho_\tau - 13050843750 \rho_2 \rho_\tau +\\
&   45805646250 \rho_1 \rho_2 \rho_\tau - 75889125 \rho_1^2 \rho_2 \rho_\tau -24218578125 \rho_2^2 \rho_\tau + 17493046875 \rho_1 \rho_2^2 \rho_\tau -\\
&   1158046875 \rho_2^3 \rho_\tau - 36356775000 \rho_\tau^2 -26724138000 \rho_1 \rho_\tau^2 - 4004587050 \rho_1^2 \rho_\tau^2 -\\
&   97864200000 \rho_2 \rho_\tau^2 - 22359847500 \rho_1 \rho_2 \rho_\tau^2 -39783656250 \rho_2^2 \rho_\tau^2 + 25721797500 \rho_\tau^3 +\\
&   3651097500 \rho_1 \rho_\tau^3 + 11282287500 \rho_2 \rho_\tau^3 + 927855000 \rho_\tau^4).
\end{split}
\]

%%%%%%%%%%%%%%%%%%%%%%%%%%%%%%%%%%%%%%%%%%%%%%%%%%%%%%%%%%%%%%%%%%%%%%%%%%%%%%%
%%%%%%%%%%%%%%%%%%%%%%%%%%%%%%%%%%%%%%%%%%%%%%%%%%%%%%%%%%%%%%%%%%%%%%%%%%%%%%%

\section{Reduction of dimensionful parameters in the MSSM}
\label{sec:red-dim1}

We move now to the dimension-1 parameters of the SSB Lagrangian, namely
the trilinear couplings $h_{t,b,\tau}$ of the SSB Lagrangian,
\refeq{SSB_L}. Again,
following the pattern in the Yukawa reduction,
in the first stage we reduce $h_{t,b}$, while $h_\tau$ will be
treated as a correction.
\[
h_i=c_i Y_i M_3 = c_i G_i M_3 g_3,\qquad i=t,b,
\]
where $M_3$ is the gluino mass.
Using the RGE for the two $h$ we get
\[
c_t=c_b=-1,
\]
where we have also used the 1-loop relation between the gaugino mass and the gauge coupling RGE
\[
2M_i\frac {dg_i}{dt}=g_i\frac {dM_i}{dt},\qquad i=1,2,3.
\]
Adding the other two gauge couplings as well as the tau Yukawa $h_\tau$
  as correction we get
\[
c_t=-\frac{A_A A_{bb} + A_{tb} B_B}{A_{bt} A_{tb} - A_{bb} A_{tt}},\qquad
c_b=-\frac{A_A A_{bt} + A_{tt} B_B}{A_{bt} A_{tb} - A_{bb} A_{tt}},
\]
where
\be
\label{rhtau}
\begin{split}
A_{tt}& = G_b^2 - \frac{16}3 - 3\rho_2 - \frac{13}{15}\rho_1,\quad
A_A = \frac{16}3 + 3\rho_2^2 + \frac{13}{15}\rho_1^2\\
A_{bb} &= G_t^2 + \rho_\tau - \frac{16}3 - 3\rho_2 - \frac 7{15}\rho_1,\quad
B_B = \frac{16}3 + 3\rho_2^2 + \frac7{15}\rho_1^2 + \rho_{h_\tau}\rho_\tau^{1/2}\\
A_{tb} &= G_b^2,\quad A_{bt} = G_t^2,\quad \rho_{h_\tau}=\frac{h_\tau}{g_3 M_3}.
\end{split}
\ee
Finally we consider the soft squared masses $m^2_\phi$ of the SSB
Lagrangian. Their reduction, according to the discussion in Section 3, takes
the form
\[
m_i^2=c_i M_3^2,\quad i=Q,u,d,H_u,H_d.
\]
The 1-loop RGE for the scalar masses reduce to the following algebraic system (where we have added
the corrections from the two gauge couplings, the tau Yukawa and $h_\tau$)
\[
\begin{split}
-12c_Q&=X_t+X_b-\frac{32}3-6\rho_2^3-\frac 2{15}\rho_1^3+\frac 15\rho_1 S,\\
-12c_u&=2X_t-\frac{32}3-\frac{32}{15}\rho_1^3-\frac 45\rho_1 S,\\
-12c_d&=2X_b-\frac{32}3-\frac 8{15}\rho_1^3+\frac 25\rho_1 S,\\
-12c_{H_u}&=3X_t-6\rho_2^3-\frac 65\rho_1^3+\frac 35\rho_1 S,\\
-12c_{H_d}&=3X_b+X_\tau-6\rho_2^3-\frac 65\rho_1^3-\frac 35\rho_1 S,
\end{split}
\]
where
\[
\begin{split}
X_t&=2G_t^2\left(c_{H_u}+c_Q+c_u\right)+2c_t^2G_t^2,\\
X_b&=2G_b^2\left(c_{H_d}+c_Q+c_d\right)+2c_b^2G_b^2,\\
X_\tau&=2\rho_\tau c_{H_d}+2\rho_{h_\tau}^2,\\
S&=c_{H_u}-c_{H_d}+c_Q-2c_u+c_d.
\end{split}
\]
Solving the above system for the coefficients $c_{Q,u,d,H_u,H_d}$ we get
\[
\begin{split}
c_Q=&-\frac{c_{Q{\rm Num}}}{D_m},\quad
c_u=-\frac 13\frac{c_{u{\rm Num}}}{D_m},\quad
c_d=-\frac{c_{d{\rm Num}}}{D_m},\\
c_{H_u}=&-\frac 23\frac{c_{Hu{\rm Num}}}{D_m},\quad
c_{H_d}=-\frac{c_{Hd{\rm Num}}}{D_m},
\end{split}
\]
where
\[
\begin{split}
D_m=&
4 (6480 + 6480 G_b^2 + 6480 G_t^2 + 6300 G_b^2 G_t^2 +
\rho_1(1836  + 1836 G_b^2  + 1836 G_t^2  +1785 G_b^2 G_t^2 )+ \\
&     \rho_\tau \left[1080  + 540 G_b^2  + 1080 G_t^2  + 510 G_b^2 G_t^2  +
 252 \rho_1  + 99 G_b^2 \rho_1  +252 G_t^2 \rho_1  + 92 G_b^2 G_t^2 \rho_1 \right]),
\end{split}
\]

\[
\begin{split}
c_{Q{\rm Num}}=&
2160 F_Q +
G_b^2(- 360 F_d - 360 F_{H_d}  + 1800 F_Q) +
G_t^2(- 360 F_{H_u} + 1800 F_Q  - 360 F_u)+\\
&G_b^2 G_t^2(- 300 F_d - 300 F_{H_d} - 300 F_{H_u} + 1500 F_Q  - 300 F_u ) +\\
&\rho_1(- 36 F_d +   36 F_{H_d}  - 36 F_{H_u}  + 576 F_Q  + 72 F_u  )+\\
&G_b^2 \rho_1(- 138 F_d  -66 F_{H_d} - 36 F_{H_u}  + 474 F_Q  + 72 F_u ) +\\
&G_t^2 \rho_1(- 36 F_d + 36 F_{H_d}  - 138 F_{H_u}  + 474 F_Q  - 30 F_u) +\\
&G_b^2 G_t^2 \rho_1(- 120 F_d - 50 F_{H_d}  -  120 F_{H_u}  + 390 F_Q  - 15 F_u ) +\\
&\rho_\tau\left[
360 F_Q  +G_b^2 (- 60 F_d + 120 F_Q )+
G_t^2 (- 60 F_{H_u} + 300 F_Q  - 60 F_u )+
\right.
\\
&G_b^2 G_t^2 (- 50 F_d - 20 F_{H_u}  + 100 F_Q  - 20 F_u )+
\rho_1 (-6 F_d - 6 F_{H_u}  + 78 F_Q  + 12 F_u )+\\
&G_b^2 \rho_1 ( - 11 F_d  + 22 F_Q ) +
G_t^2 \rho_1 ( - 6 F_d  -20 F_{H_u}  + 64 F_Q  - 2 F_u ) +\\
&\left.
G_b^2 G_t^2 \rho_1 ( -9 F_d  - 4 F_{H_u}  +18 F_Q  - 3 F_u )
\right],
\end{split}
\]

\[
\begin{split}
c_{u{\rm Num}}=&
6480 F_u + 6480 F_u G_b^2 +
G_t^2(- 2160 F_{H_u}  - 2160 F_Q  + 4320 F_u ) +\\
& G_b^2 G_t^2(  360 F_d  + 360 F_{H_d}  - 2160 F_{H_u}  -1800 F_Q  + 4140 F_u )+\\
&    \rho_1( 432 F_d  - 432 F_{H_d} +432 F_{H_u}  + 432 F_Q  + 972 F_u )+\\
&    G_b^2 \rho_1( 432 F_d  -432 F_{H_d}  + 432 F_{H_u}  + 432 F_Q  + 972 F_u )+\\
&   G_t^2 \rho_1(432 F_d  - 432 F_{H_d}  - 180 F_{H_u}  - 180 F_Q  +360 F_u )+\\
&    G_b^2 G_t^2 \rho_1( 522 F_d  - 318 F_{H_d}  - 192 F_{H_u}  - 90 F_Q  + 333 F_u ) +\\
&   \rho_\tau \left[
     1080 F_u  +
     540 G_b^2 F_u + G_t^2(  - 360 F_{H_u}   - 360 F_Q   + 720 F_u )+ \right.\\
&      G_b^2 G_t^2 (60 F_d   - 180 F_{H_u}   - 120 F_Q   + 330 F_u )  +
  \rho_1( 72 F_d   + 72 F_{H_u}   + 72 F_Q   + 108 F_u )  +\\
&   G_b^2 \rho_1( 36 F_{H_u}   + 27 F_u ) +
72 G_t^2 \rho_1 (F_d   - 12 F_{H_u}   - 12 F_Q   + 24 F_u )  +\\
&   \left.
G_b^2 G_t^2 \rho_1 (9 F_d   + 4 F_{H_u}   - 18 F_Q   + 3 F_u )
    \right],
\end{split}
\]

\[
\begin{split}
c_{d{\rm Num}}=&
2160 F_d + G_b^2(1440 F_d  - 720 F_{H_d}  - 720 F_Q ) + 2160 F_d G_t^2 +\\
&   G_b^2 G_t^2 (1380 F_d  - 720 F_{H_d}  + 120 F_{H_u}  -  600 F_Q  + 120 F_u ) +\\
&   \rho_1( 540 F_d  + 72 F_{H_d}  - 72 F_{H_u}  - 72 F_Q  + 144 F_u )+\\
&    G_b^2 \rho_1( 336 F_d  -132 F_{H_d}  - 72 F_{H_u}  - 276 F_Q  + 144 F_u ) +\\
&   G_t^2 \rho_1( 540 F_d  + 72 F_{H_d}  - 72 F_{H_u}  - 72 F_Q  + 144 F_u )+\\
&   G_b^2 G_t^2 \rho_1( 321 F_d  - 134 F_{H_d}  -36 F_{H_u}  - 240 F_Q  + 174 F_u ) +\\
&   \rho_\tau\left[
     360 F_d  + G_b^2( 60 F_d   - 120 F_Q )  + 360 F_d G_t^2  +
   G_b^2 G_t^2(50 F_d   + 20 F_{H_u}   - 100 F_Q   +20 F_u )+ \right.\\
&     \rho_1(72 F_d   - 12 F_{H_u}  -12 F_Q   + 24 F_u)   +  G_b^2 \rho_1 ( 11 F_d -22 F_Q)    +\\
&   \left.  G_t^2 \rho_1( 72 F_d   - 12 F_{H_u}   -12 F_Q   + 24 F_u )+
     G_b^2 G_t^2 \rho_1(9 F_d   + 4 F_{H_u}  - 18 F_Q   + 3 F_u )
    \right],
\end{split}
\]

\[
\begin{split}
c_{Hu{\rm Num}}=&
3240 F_{H_u} + 3240 F_{H_u} G_b^2 + G_t^2( 1620 F_{H_u}  - 1620 F_Q  - 1620 F_u )+\\
&    G_b^2 G_t^2( 270 F_d  + 270 F_{H_d}  + 1530 F_{H_u}  - 1350 F_Q  - 1620 F_u )+ \\
&    \rho_1(- 162 F_d  + 162 F_{H_d}  + 756 F_{H_u}  - 162 F_Q  + 324 F_u )+\\
&    G_b^2 \rho_1(- 162 F_d  + 162 F_{H_d}  + 756 F_{H_u}  - 162 F_Q  + 324 F_u )+\\
&   G_t^2 \rho_1(-162 F_d  + 162 F_{H_d}  + 297 F_{H_u}  - 621 F_Q  - 135 F_u )+\\
&   G_b^2 G_t^2 \rho_1 (- 81 F_d  + 234 F_{H_d}  + 276 F_{H_u}  - 540 F_Q  - 144 F_u ) +\\
&  \rho_\tau \left[
  540 F_{H_u}  + 270 F_{H_u} G_b^2   + G_t^2(270 F_{H_u}   - 270 F_Q    - 270 F_u )+ \right.\\
&       G_b^2 G_t^2( 45 F_d    + 120 F_{H_u}    - 90 F_Q    - 135 F_u )+
   \rho_1(-27 F_d    + 99 F_{H_u}    - 27 F_Q    + 54 F_u )   +\\
&   G_b^2 \rho_1( 36 F_{H_u}    + 27 F_u    - 27 F_d )   + G_t^2 \rho_1( 36 F_{H_u}    - 90 F_Q    - 9 F_u )   +\\
&  \left.  G_b^2 G_t^2 \rho_1( 9 F_d    + 4 F_{H_u}    - 18 F_Q    + 3 F_u )
\right],
\end{split}
\]

\[
\begin{split}
c_{Hd{\rm Num}}=&
2160 F_{H_d}+ G_b^2 (- 1080 F_d + 1080 F_{H_d}  - 1080 F_Q ) + 2160 F_{H_d} G_t^2+\qquad\qquad\qquad\qquad\\
&    G_b^2 G_t^2(- 1080 F_d  + 1020 F_{H_d}  +180 F_{H_u}  - 900 F_Q  + 180 F_u )+\\
&    \rho_1( 108 F_d  +504 F_{H_d} + 108 F_{H_u}  + 108 F_Q  - 216 F_u )+\\
&    G_b^2 \rho_1(- 198 F_d  +198 F_{H_d}  + 108 F_{H_u}  - 198 F_Q  - 216 F_u ) +\\
&   G_t^2 \rho_(108 F_d 1 + 504 F_{H_d}  + 108 F_{H_u}  + 108 F_Q  -216 F_u )+\\
&    G_b^2 G_t^2 \rho_1(- 201 F_d  + 184 F_{H_d}  +156 F_{H_u}  - 150 F_Q  - 159 F_u )
\end{split}
\]
and
\[
\begin{split}
F_Q &= 2 c_t^2 G_t^2 + 2 c_b^2 G_b^2 - \frac{32}{3} - 6 \rho_2^3 - \frac{2}{15} \rho_1^3,\\
F_u &= 4 c_t^2 G_t^2 - \frac{32}{3} - \frac{32}{15} \rho_1^3,\\
F_d &= 4 c_b^2 G_b^2 - \frac{32}{3} - \frac{8}{15} \rho_1^3,\\
F_{H_u}& = 6 c_t^2 G_t^2 - 6 \rho_2^3 - \frac{6}{5} \rho_1^3,\\
F_{H_d}& = 6 c_b^2 G_b^2 + 2 \rho_{h_\tau}^2 - 6 \rho_2^3 - \frac{6}{5} \rho_1^3,
\end{split}
\]
while $G_{t,b}^2$, $\rho_{1,2,\tau}$ and $\rho_{h_\tau}$ has been defined in Eqs.(\ref{Gt2_Gb2},\ref{r1_r2_rtau},\ref{rhtau})
respectively.
For our completely reduced system, i.e. $g_3,Y_t,Y_b,h_t,h_b$, the coefficients of the soft masses
become
\[
c_Q=c_u=c_d=\frac 23,\quad c_{H_u}=c_{H_d}=-1/3,
\]
obeying the celebrated sum rules
\[
\frac{m_Q^2+m_u^2+m_{H_u}^2}{M_3^2}=c_Q+c_u+c_{H_u}=1,\qquad
\frac{m_Q^2+m_d^2+m_{H_d}^2}{M_3^2}=c_Q+c_d+c_{H_d}=1.
\]

\bigskip
The $\mu$ parameter of the superpotential cannot be reduced, at least in a
simple way of the form $\mu=c_\mu M_3 g_3$ as an ansatz at one loop.
The parameter $m_3^2$ in the SSB sector could in principle be reduced in
favor of $\mu$ and $M_3$, but in our analysis we keep
$m_3^2$ as independent parameter.
However, it should be noted that the requirement of radiative electroweak
symmetry breaking
  (EWSB) relates $\mu$ and $m_3^2$, and leaves only one of them as an
  independent parameter, which we choose to be $\mu$.

%%%%%%%%%%%%%%%%%%%%%%%%%%%%%%%%%%%%%%%%%%%%%%%%%%%%%%%%%%%%%%%%%%%%%%%%%%%%%%%
%%%%%%%%%%%%%%%%%%%%%%%%%%%%%%%%%%%%%%%%%%%%%%%%%%%%%%%%%%%%%%%%%%%%%%%%%%%%%%%

\section{Phenomenological constraints}
\label{sec:constraints}

In this section we will briefly describe the phenomenological
constraints that we apply on the parameter space of the reduced MSSM, as
described above.

%%%%%%%%%%%%%%%%%%%%%%%%%%%%%%%%%%%%%%%%%%%%%%%%%%%%%%%%%%%%%%%%%%%%%%%%%%%%%%%

\subsection{Flavor constraints}

As  additional constraints we consider four types of flavor contraints,
where SUSY is know to have a possible impact. We consider
the flavour observables $\br(b \to s \ga)$, $\br(B_s \to \mu^+ \mu^-)$,
$\br(B_u \to \tau \nu)$ and $\Delta B_{M_s}$.%
\footnote{We do not employ the very latest experimental data, but this
  has a minor impact on our analysis.}%
~The uncertainties are the linear combination of the experimental error and
twice the theoretical uncertainty in the MSSM (if no specific MSSM estimate is
avialabe we use the SM uncertainty).

For the branching ratio $\br(b \to s \gamma)$, we take the value
given by the Heavy Flavour Averaging Group (HFAG) is~\cite{bsgth,HFAG}
\be
\frac{\br(b \to s \ga)^{\rm exp}}{\br(b \to s \ga)^{\SM}} = 1.089 \pm 0.27~.
\label{bsgaexp}
\ee
For the branching ratio $\br(B_s \to \mu^+ \mu^-)$ we use a combination of CMS
and LHCb data~\cite{Bobeth:2013uxa,RmmMFV,LHCbBsmm,CMSBsmm,BsmmComb}
\be
\br(B_s \to \mu^+ \mu^-) = (2.9 \pm 1.4) \times 10^{-9}~.
\ee
For the $B_u$ decay to $\tau \nu$ we use the
limit~\cite{SuFla,HFAG,PDG14}
\be
\frac{\br(B_u \to \tau\nu)^{\rm exp}}{\br(B_u \to \tau\nu)^{\SM}} =
                                                          1.39 \pm 0.69~.
\ee
As our final flavor observalbe we include $\De M_{B_s}$
as~\cite{Buras:2000qz,Aaij:2013mpa}
\be
\frac{\De M_{B_s}^{\rm exp}}{\De M_{B_s}^{\SM}} = 0.97 \pm 0.2~.
\ee
Our theory evaluations are obtained with the code
\texttt{SuFla}~\cite{SuFla}.

\medskip
We do not include a bound from the cold dark matter (CDM) density.
It is well known that the lightest neutralino, being the lightest
supersymmetric particle (LSP) in our model, is an
excellent candidate for CDM~\cite{EHNOS}.
However, the models could easily be extended to contain (a) small R-parity
violating term(s)~~\cite{herbi,herbi2,herbi3,herbi4}. They
would have a small impact on the collider phenomenology discussed here
(apart from the fact that the SUSY search strategies could not rely on a
`missing energy' signature), but would remove the CDM bound
completely. Other mechanisms, not
involving R-parity violation (and keeping the `missing energy'
signature), that could be invoked if the amount of CDM appears to be
too large, concern the cosmology of the early universe.  For instance,
``thermal inflation''~\cite{thermalinf} or ``late time entropy
injection''~\cite{latetimeentropy} could bring the CDM density into
agreement with the WMAP measurements.  This kind of modifications of
the physics scenario neither concerns the theory basis nor the
collider phenomenology, but could have a strong impact on the CDM
derived bounds.
(Lower values than the ones permitted by the experimental measurements
are naturally allowed if another particle than the lightest neutralino
constitutes
CDM.)

We will briefly comment on the anomalous magnetic moment of the muon,
$(g-2)_\mu$, at the end of \refse{sec:numanal}.

%%%%%%%%%%%%%%%%%%%%%%%%%%%%%%%%%%%%%%%%%%%%%%%%%%%%%%%%%%%%%%%%%%%%%%%%%%%%%%%

\subsection{The light Higgs boson mass}
\label{sec:Mh}

Due to the fact that the quartic couplings in the Higgs potential are
given by the SM gauge couplings, the lightest Higgs boson mass is not a
free parameter, but predicted in terms of the other model parameters.
Higher-order corrections are crucial for a precise prediction of
$\Mh$, see \citeres{habilSH,awb2,PomssmRep} for reviews.

The spectacular discovery of a Higgs boson at ATLAS and CMS, as
announced in July 2012~\cite{:2012gk,:2012gu} can be interpreted as the
discovery of the light $\cp$-even Higgs boson of the MSSM Higgs
spectrum~\cite{Mh125} (see also \citeres{hifi,hifi2} and references
therein).
The experimental average for the (SM) Higgs boson mass is taken to
be~\cite{Aad:2015zhl}
\begin{align}
\MH^{\rm exp} &= 125.1 \pm 0.3 \gev~.
\end{align}
Adding a $3\, (2) \gev$ theory
uncertainty~\cite{Degrassi:2002fi,Buchmueller:2013psa,BHHW} for
the Higgs boson mass calculation in the MSSM we arrive at
\begin{align}
\Mh &= 125.1 \pm 3.1\, (2.1) \gev
\label{Mhexp}
\end{align}
as our allowed range.

For the lightest Higgs mass prediction we used the code
\FH~\cite{Degrassi:2002fi,BHHW,FeynHiggs} (version 2.14.0\,beta).
The evaluation of Higgs boson masses within \FH\
is based on the combination of a Feynman-diagrammatic calculation and
a resummation of the (sub)leading and logarithms contributions of the
(general) type
$\log(\mstop{}/\mt)$ in all orders of perturbation theory.
This combination ensures a reliable evaluation of $\Mh$ also for large
SUSY mass scales (see \refse{sec:numanal} below).
With respect to previous versions several refinements in the combination
of the fixed order log resummed calculation have been
included, see \citere{BHHW}. They resulted not only in a
more precise $\Mh$ evaluation for high SUSY mass scales, but in
particular in a downward shift of $\Mh$ at the level of \order{2 \gev}
for large SUSY masses.

In our previous analysis \cite{MTZ:14} the Higgs boson mass was calculated using a ``mixed-scale'' one-loop RG approach,
which captures only the leading corrections up to two-loop order.
Consequently, our new implementation of the $\Mh$ calculation is substantially more sophisticated and in particular
reliable for high stop mass scales.
Furthermore, in that previous analysis no B physics constraints  were used,
which now pose relevant constraints on the allowed parameters space and thus on the prediction of the SUSY spectrum.

%%%%%%%%%%%%%%%%%%%%%%%%%%%%%%%%%%%%%%%%%%%%%%%%%%%%%%%%%%%%%%%%%%%%%%%%%%%%%%%
%%%%%%%%%%%%%%%%%%%%%%%%%%%%%%%%%%%%%%%%%%%%%%%%%%%%%%%%%%%%%%%%%%%%%%%%%%%%%%%

\section{Numerical analysis}
\label{sec:numanal}

In this section we analyze the particle spectrum predicted by the
reduced MSSM.  So far the relations among reduced parameters in terms
of the fundamental ones derived in \refses{sec:red-dim0} and
\ref{sec:red-dim1} had a part which was RGI and a another part
originating from the corrections, which are scale dependent. In our
analysis here we choose the unification scale to apply the corrections
to the RGI relations.
It should be noted that we are
assuming a covering GUT, and thus unification of the three
gauge couplings, as well as a unified gaugino mass
$M$  at that scale. Also to be noted is that in the dimensionless sector of the
theory since $Y_\tau$ cannnot be reduced in favor of the fundamental
parameter $\al_3$, the mass of the $\tau$ lepton is an input parameter and
consequently $\rho_\tau$, is an independent parameter too.  At low
energies, we fix the values of $\rho_{\tau}$ and $\tan\beta$ using
the mass of the tau lepton $m_{\tau}(M_Z)$.  For each value of
$\rho_{\tau}$ there is a corresponding value of $\tan\beta$ that
gives the appropriate $m_{\tau}(M_Z)$.  Then we use the value found
for $\tan\beta$ together with $G_{t,b}$, as obtained from the
reduction equations and their respective corrections, to determine the top and bottom quark masses.  We
require that both the bottom and top masses are within 2$\sigma$ of
their experimental value, which singles out  large $\tan\beta$
values, $\tan\beta \sim 42 - 47$.
Correspondingly, in the dimensionful sector of the theory the $\rho_{h_\tau}$
is a free parameter, since $h_\tau$ cannot be reduced in favor of the
fundamental parameter $M$ (the unified gaugino mass scale).
  $\mu$ is a free parameter, as it cannot be reduced in favor of $M_3$
as discussed above. On the other hand $m_3^2$ could be reduced, but here
it is chosen to leave it free.
However, $\mu$ and $m_3^2$ are restricted from the requirement of
EWSB, and only $\mu$ is taken as an independent parameter.
Finally, the other parameter in the Higgs-boson sector, the $\cp$-odd
Higgs-boson mass $\MA$ is evaluated from $\mu$, as well as from $m_{H_u}^2$ and
$m_{H_d}^2$, which are obtained from the reduction equations.
In total we vary the parameters $\rho_\tau$, $\rho_{h_\tau}$, $M$
  and $\mu$.

\begin{figure}[!]
\begin{center}
\includegraphics[width=0.78\textwidth,height=8cm]{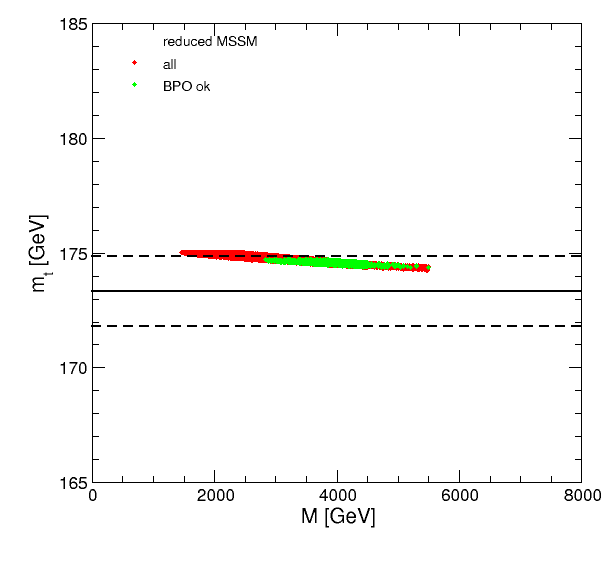}\\[0.1em]
\includegraphics[width=0.78\textwidth,height=8cm]{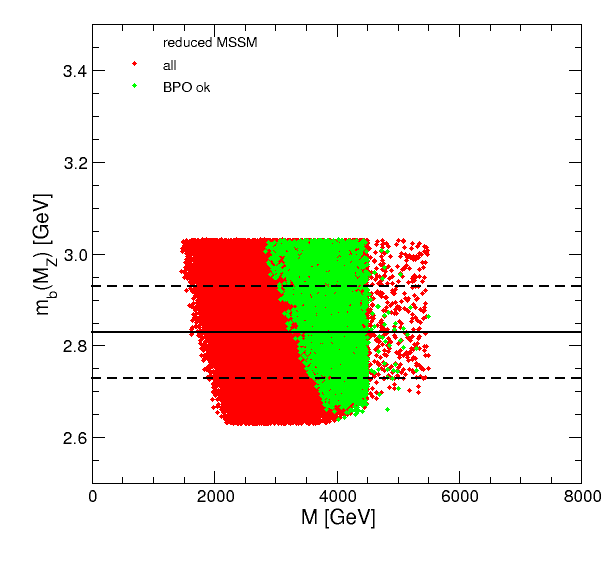}\\[-1em]
\caption{The upper (lower) plot shows our results within the reduced
    MSSM for the top (bottom) quark mass. The horizontal lines indicate the
    experimental values as given in \refeq{mtmb}.
}
\label{fig:mf}
\end{center}
\vspace{-1em}
\end{figure}

We start our numerical analysis with the top and the bottom quark masses.
As mentioned above, the variation of $\rho_\tau$ yields the
values of $\mt$ (the top pole mass) and $\mb(\MZ)$, the running
bottom quark mass at the $Z$~boson mass scale,  where
  scan points which are not within $2\sigma$ of
the experimental data are neglected. This is shown in
\reffi{fig:mf}. The experimental values are indicated by the horizontal lines
and are taken to be~\cite{PDG14},
\be
\mt = 173.34 \pm 1.52 \gev~, \quad
\mb(\MZ) = 2.83 \pm 0.1 \gev~,
\label{mtmb}
\ee
with the uncertainties at the $2\,\sig$~level. One can see that the
scan yields many parameter points that
are in very good agreement with the experimental data.

%%%%%%%%%%%%%%%%%%%%%%%%% F I G U R E %%%%%%%%%%%%%%%%%%%%%%%%%%%%%%%%%%%%%%%%%
%%%%%%%%%%%%%%%%%%%%%%%%% F I G U R E %%%%%%%%%%%%%%%%%%%%%%%%%%%%%%%%%%%%%%%%%

\begin{figure}[!]
\mbox{}\vspace{-5em}
\begin{center}
\includegraphics[width=0.90\textwidth]{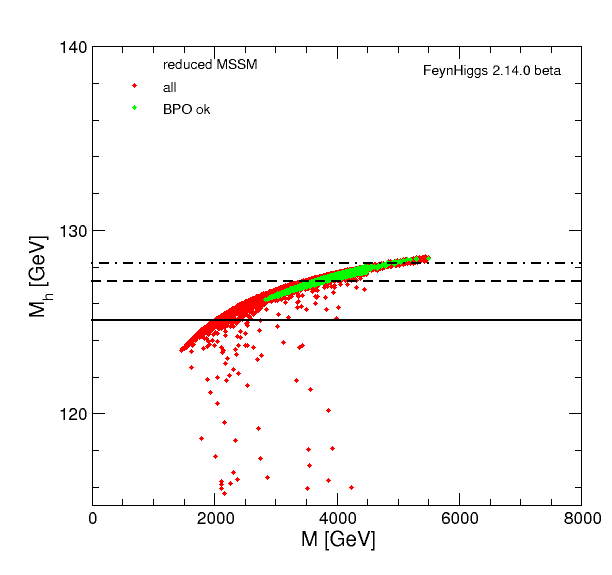}
\caption{The lightest Higgs boson mass, $\Mh$, as a function of $M$
(the common gaugino mass at the unification scale)
in the reduced MSSM. The red points is the full model prediction. The
  green points fulfill the $B$-physics constraints (see text).
}
\label{fig:Mh}
\end{center}
%\vspace{-1em}
\end{figure}

\medskip
We continue our numerical investigation with the analysis of the
lightest MSSM Higgs-boson mass.
The prediction for $\Mh$ is shown in
\reffi{fig:Mh} as a function of $M$ (the common gaugino mass at the
unification scale) in the range
$1 \tev \lsim M \lsim 6 \tev$. The lightest Higgs mass ranges in
\be
\Mh \sim 124-129 \gev~ ,
\label{eq:Mhpred}
\ee
where we discard the ``spreaded'' points with possibly lower masses,
which result from a numerical instability in the Higgs-boson mass
calculation.
One should keep in mind that these predictions are subject to a theory
uncertainty of $3 (2) \gev$, see above.
The red points correspond to the full parameter scan, whereas the green
points are the subset that is in agreement with the $B$-physics observables
as discussed above (which do not exhibit any numerical instability). The
inclusion of the flavor observables shifts the lower bound for $\Mh$ up to
$\sim 126 \gev$.

The horizontal lines in \reffi{fig:Mh} show the central value of the
experimental measurement (solid), the $\pm 2.1 \gev$ uncertainty (dashed) and
the $\pm 3.1 \gev$ uncertainty (dot-dashed). The requirement to obtain a
light Higgs boson mass value in the correct range yields an upper limit
on $M$ of about $5\, (4) \tev$ for $\Mh = 125.1 \pm 2.1\, (3.1) \gev$.

%%%%%%%%%%%%%%%%%%%%%%%%% F I G U R E %%%%%%%%%%%%%%%%%%%%%%%%%%%%%%%%%%%%%%%%%
%%%%%%%%%%%%%%%%%%%%%%%%% F I G U R E %%%%%%%%%%%%%%%%%%%%%%%%%%%%%%%%%%%%%%%%%

%%%%%%%%%%%%%%%%%%%%%%%%% F I G U R E %%%%%%%%%%%%%%%%%%%%%%%%%%%%%%%%%%%%%%%%%
\begin{figure}[!]
\begin{center}
\includegraphics[width=0.78\textwidth,height=8cm]{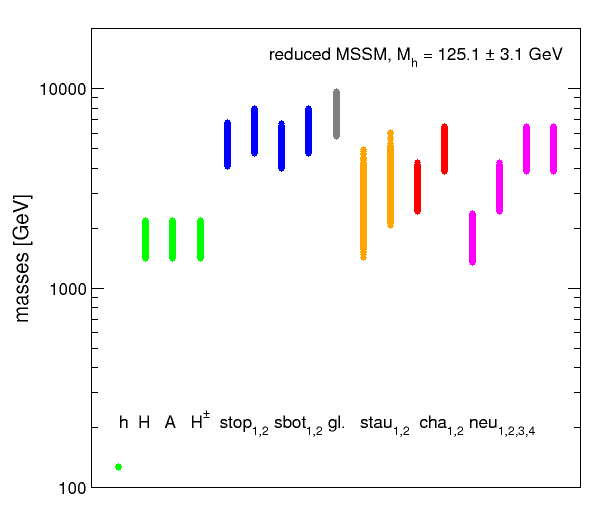}\\[0.1em]
\includegraphics[width=0.78\textwidth,height=8cm]{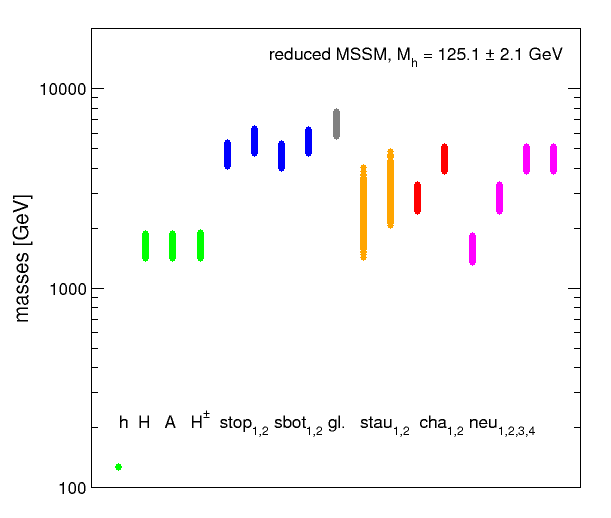}\\[-1em]
\caption{The upper (lower) plot shows the spectrum of the reduced MSSM
after imposing the constraint
$\Mh = 125.1 \pm 3.1\,(2.1) \gev$.
The points shown are in agreement with the $B$-physics observables.
The light (green) points on the left are the various Higgs
boson masses. The dark (blue) points following are the
two scalar top and bottom masses, followed by the lighter
(gray) gluino mass. Next come the lighter (beige) scalar
tau masses. The darker (red) points to the right are the
two chargino masses followed by the lighter shaded (pink)
points indicating the neutralino masses.
}
\label{fig:masses}
\end{center}
\vspace{-1em}
\end{figure}
%%%%%%%%%%%%%%%%%%%%%%%%% F I G U R E %%%%%%%%%%%%%%%%%%%%%%%%%%%%%%%%%%%%%%%%%

\medskip
Naturally the $\Mh$ limit also sets an upper limit on the low-energy
SUSY masses. The full particle spectrum of the reduced MSSM
(where we restricted ourselves as before to the third generation of
sfermions)
compliant with the $B$-physics observables
is shown in \reffi{fig:masses}.  In the upper (lower) plot we impose
$\Mh = 125.1 \pm 3.1\, (2.1) \gev$.  Including
the Higgs mass constraints in general favors the somewhat higher part of the
SUSY particle mass spectra. The tighter $\Mh$ range cuts off the very
high SUSY mass scales.
The lighter SUSY particles are given by the electroweak spectrum, which
starts around $\sim 1.3 \tev$. They will mostly remain unobservable at the LHC
and at future $e^+e^-$ colliders such as the ILC or CLIC, with only the
very lower range mass range below $\sim 1.5 \tev$ might be observable at
CLIC (with $\sqrt{s} = 3 \tev$). The colored mass spectrum starts at
around $\sim 4 \tev$, which will remain unobservable at the
(HL-)LHC. However, the colored spectrum would be accessible at the
FCC-hh~\cite{fcc-hh}. The same applies to the heavy Higgs-boson
spectrum. The four ``new'' Higgs bosons will likely remain outside the
reach of the (HL-)LHC, ILC and CLIC, again with the very lower part of
the spectrum potentially accessible at CLIC. However, the full Higgs
boson spectrum would be covered at the FCC-hh~\cite{fcc-hh}.

%%%%%%%%%%%%%%%%%%%%% T A B L E %%%%%%%%%%%%%%%%%%%%%%%%%%%%%%%%%%%%%%%%%%%%%%
\begin{table}[t!]
\renewcommand{\arraystretch}{1.5}
\centering
\begin{tabular}{|c|rrrrrrrrr|}
\hline
& $\Mh$ & $\MH$ & $\MA$ & $\MHp$ & $\mstop1$ & $\mstop2$ &
  $\msbot1$& $\msbot2$ & $\mgl$ \\
\hline
light        & 126.2 & 1433 & 1433 & 1446 & 4052 & 4736 & 3989 & 4723 & 5789 \\
$\de\Mh=2.1$ & 127.2 & 1570 & 1570 & 1572 & 5361 & 6289 & 5282 & 6279 & 7699 \\
$\de\Mh=3.1$ & 128.1 & 1886 & 1886 & 1888 & 6762 & 7951 & 6653 & 7943 & 9683 \\
\hline
\end{tabular}

\vspace{1em}
\begin{tabular}{|c|rrrrrrrrr|}
\hline
& $\mstau1$ & $\mstau2$ &
  $\mcha1$ & $\mcha2$ & $\mneu1$ & $\mneu2$ & $\mneu3$ & $\mneu4$ & $\tb$ \\
\hline
light        & 1906 & 2066 & 2430 & 3867 & 1339 & 2430 & 3864 & 3866 & 42.6 \\
$\de\Mh=2.1$ & 1937 & 2531 & 3299 & 5166 & 1833 & 3299 & 5114 & 5116 & 43.1 \\
$\de\Mh=3.1$ & 3153 & 3490 & 4248 & 6464 & 2376 & 4248 & 6462 & 6464 & 45.2 \\
\hline
\end{tabular}
\caption{
Three example spectra of the reduced MSSM. ``light'' has the smallest $\neu1$
in our sample, ``$\de\Mh = 2.1 (3.1)$'' has the largest $\mneu1$ for
$\Mh \le 125.1 + 2.1 (3.1) \gev$.
All masses are in GeV and rounded to 1 (0.1)~GeV (for the light Higgs mass).
}
\label{tab:spectrum}
\renewcommand{\arraystretch}{1.0}
\end{table}
%%%%%%%%%%%%%%%%%%%%% T A B L E %%%%%%%%%%%%%%%%%%%%%%%%%%%%%%%%%%%%%%%%%%%%%%

In \refta{tab:spectrum} we show three example spectra of the reduced MSSM,
which span the mass range of the parameter space that is in agreement with the
$B$-physics observables and the Higgs-boson mass measurement.
The four Higgs boson masses are denoted as $\Mh$, $\MH$, $\MA$ and
  $\MHp$. $\mstop{1,2}$, $\msbot{1,2}$, $\mgl$, $\mstau{1,2}$,
  are the scalar top, scalar bottom, gluino and scalar tau
  masses, respectively. $\mcha{1,2}$ and $\mneu{1,2,3,4}$ denote the
  chargino and neutralino masses.
The rows labelled ``light'' correspond to
the spectrum with the smallest $\mneu1$ value (which is independent of upper
limit in $\Mh$). This point is an example for the lowest $\Mh$ values that we
can reach in our scan. As discussed above, the heavy Higgs boson spectrum starts
above $1.4 \tev$, which is at the borderline of the reach of CLIC with
$\sqrt{s} = 3 \tev$. The colored spectrum is found between $\sim 4 \tev$ and
$\sim 6 \tev$, outside the range of the (HL-)LHC. The LSP has a mass of
$\mneu1 = 1339$, which might offer the possibility of
$e^+e^- \to \neu1 \neu1 \ga$ at CLIC. All other electroweak particles are too
heavy to be produced at CLIC or the (HL-)LHC.
``$\de\Mh = 2.1 (3.1)$'' has the largest $\mneu1$ for
$\Mh \le 125.1 + 2.1 (3.1) \gev$. While, following the mass relations in the
reduced MSSM, the mass spectra are substantially heavier than in the ``light''
case, one can also observe that the smaller upper limit on $\Mh$ results in
substantially lower upper limits on the various SUSY and Higgs-boson
masses. However, even in the case of $\de\Mh = 2.1 \gev$, all particles are
outside the reach of the (HL-)LHC and CLIC. On the other hand, all spectra
offer good possibilities for their discovery at the FCC-hh~\cite{fcc-hh}, as
discussed above.

\bigskip
Finally, we note that with such a heavy SUSY spectrum, despite the large
values of $\tb$,
the anomalous magnetic moment of the muon, $(g-2)_\mu$
(with $a_\mu \equiv (g-2)_\mu/2$), gives only a negligible correction to
the SM prediction.
The comparison of the experimental result and the SM value shows a
deviation of $\sim 3.5\,\sig$~\cite{newBNL,g-2,Jegerlehner}.  Consequently,
since the results would be very close to the SM results, the
model has the same level of difficulty with the $a_\mu$ measurement as
the SM.

\bigskip
To summarize, the reduced MSSM naturally results in a light Higgs boson
in the mass range measured at the LHC. On the other hand, the rest of
the spectrum will remain (likely) unaccessible at the (HL-)LHC, ILC and
CLIC, where such a heavy spectrum also results in SM-like light Higgs
boson, in agreement with LHC measurements~\cite{HcoupLHCcomb}. In other
words, the model is naturally in full agreement with all LHC
measurements. It can be tested definitely at the FCC-hh, where large
parts of the spectrum would be in the kinematic reach.

%%%%%%%%%%%%%%%%%%%%%%%%%%%%%%%%%%%%%%%%%%%%%%%%%%%%%%%%%%%%%%%%%%%%%%%%%%%%%%%
%%%%%%%%%%%%%%%%%%%%%%%%%%%%%%%%%%%%%%%%%%%%%%%%%%%%%%%%%%%%%%%%%%%%%%%%%%%%%%%

\section{Conclusions}
\label{sec:conclusions}

In the present paper we have examined the reduced MSSM, in which we first
calculate the exact relations among soft scalar and gaugino masses at the
unification scale. This constitutes an interesting improvement
w.r.t.\ previous analyses~\cite{MTZ:14}, which relied
on the existence of a ``sum rule'' among soft scalar and gaugino masses,
where due to the ``simple'' nature of the constraint agreement with
  experimental data could be realized more easily.
It should be noted that in the reduced MSSM
the ``sum rule'' still is valid. However, here we
have the exact relations among these masses, and consequently the
dimensionful SSB mass relations are as those among the dimensionless
couplings.

In our phenomenological analysis we have derived the spectrum of the
  reduced MSSM as a function of the common gaugino mass at the GUT scale.
The light Higgs boson mass was evaluated with the latest (preliminary) version
of \FH~\cite{BHHW}, which yields more reliable results in the case of very
large SUSY mass scales, as it turns out to be the case in our analysis.
The resulting spectrum was confronted with various $B$-physics constraints.
We find that the lightest Higgs mass is in very good agreement with the
measured value and its experimental and theoretical uncertainties. The SUSY
Higgs boson mass scale is found above $\sim 1.3 \tev$, rendering the light
MSSM Higgs boson SM-like, in perfect agreement with the experimental data.
The electroweak SUSY spectrum starts at 1.3~TeV and the colored spectrum at
$\sim$4~TeV. Consequently, the reduced MSSM is in natural agreement with all
LHC measurements and searches. The SUSY and heavy Higgs particles will
likely escape the detection at the LHC, as well as at ILC and CLIC.
On the other hand, the FCC-hh will be able to fully test the predicted
parameter space.

%%%%%%%%%%%%%%%%%%%%%%%%%%%%%%%%%%%%%%%%%%%%%%%%%%%%%%%%%%%%%%%%%%%%%%%%%%%%%%%
%%%%%%%%%%%%%%%%%%%%%%%%%%%%%%%%%%%%%%%%%%%%%%%%%%%%%%%%%%%%%%%%%%%%%%%%%%%%%%%

\subsection*{Acknowledgements}

\begingroup \small
We thank
H.~Bahl,
T.~Hahn,
W.~Hollik,
D.~L\"ust
and
E.~Seiler
for helpful discussions.
The work of S.H.\ is supported in
part by the MEINCOP Spain under contract FPA2016-78022-P, in part by the
Spanish Agencia Estatal de Investigaci{\' o}n (AEI) and the EU Fondo
Europeo de Desarrollo Regional (FEDER) through the project
FPA2016-78645-P, and in part by the AEI through the grant IFT Centro de
Excelencia Severo Ochoa SEV-2016-0597.
The work of M.M.\ is supported partly by  UNAM PAPIIT grant
IN111518.
The work of N.T.\ and G.Z.\ are supported by the
COST actions CA15108 and CA16201.
N.T.\ and G.Z.\ thank the CERN TH Department for their hospitality.
G.Z.\ thanks the MPI Munich for hospitality and the A.v.Humboldt Foundation for support.\\
Finally, we would like to acknowledge the very constructive discussion with the referee delving us
even deeper into the problem.
\endgroup

%%%%%%%%%%%%%%%%%%%%%%%%%%%%%%%%%%%%%%%%%%%%%%%%%%%%%%%%%%%%%%%%%%%%%%%%%%%%%%%
%%%%%%%%%%%%%%%%%%%%%%%%%%%%%%%%%%%%%%%%%%%%%%%%%%%%%%%%%%%%%%%%%%%%%%%%%%%%%%%

\end{document}